\documentclass[aps,prl,twocolumn,superscriptaddress,longbibliography,floatfix]{revtex4-2}
\usepackage{hyperref}
\usepackage{graphicx}
\usepackage{physics}
\usepackage{mathrsfs}
\usepackage{amsmath}
\usepackage{amssymb}
\usepackage{amsfonts}
\usepackage{mathtools}
\usepackage{xcolor}
\usepackage[normalem]{ulem}
\usepackage{xparse}

\usepackage{color}
\usepackage{tikz}

\hypersetup{colorlinks=true,citecolor=blue,linkcolor=blue,urlcolor=blue}

\newcommand{\im}{{\rm i}}
\newcommand{\one}{{\rm I}}
\newcommand{\two}{{\rm I\hspace{-1.2pt}I}}

\begin{document} 
\title{The quantum Mpemba effect in symmetric random Clifford circuits} 
\author{Shion Yamashika}
\affiliation{Department of Engineering Science, The University of Electro-Communications, Tokyo 182-8585, Japan.}
\author{Filiberto Ares}
\affiliation{SISSA and INFN, via Bonomea 265, 34136 Trieste, Italy.}

\author{Pasquale Calabrese}
\affiliation{SISSA and INFN, via Bonomea 265, 34136 Trieste, Italy.}

\begin{abstract}
The quantum Mpemba effect is the counterintuitive phenomenon whereby a quantum state initially farther from equilibrium relaxes faster than one initially closer to it. We demonstrate this effect in random Clifford circuits with a conserved $\mathrm{U}(1)$ charge, using the entanglement asymmetry to characterize relaxation through dynamical symmetry restoration. Exact numerical simulations show that an initially more asymmetric state can become locally more symmetric than an initially less asymmetric one. We explain this behavior by mapping the dynamics onto a charge-conserving quantum automaton, in which the decay of the entanglement asymmetry is controlled by the statistics of encounters between two particle species evolving according to a symmetric simple exclusion process. This mapping yields an analytic expression for the entanglement asymmetry and provides a simple microscopic mechanism for the quantum Mpemba effect: stronger initial symmetry breaking corresponds to a denser particle configuration, which enhances the frequency of particle encounters and thereby accelerates symmetry restoration.

\end{abstract}

\maketitle

\textit{Introduction}.--- The Mpemba effect is a genuinely nonequilibrium phenomenon that defies the intuition built from quasistatic processes: a state that starts farther from equilibrium can relax faster than one that starts closer to equilibrium~\cite{teza2025}. Named after the counterintuitive observation that hotter water can sometimes freeze faster than colder water~\cite{mpemba1969,bechhoefer2021}, the effect was recently extended to quantum systems, where it is known as the quantum Mpemba effect (QME)~\cite{ares2025}.

The QME has been studied in both closed~\cite{ares2023,Rylands2024,murciano2024,chalas2024,yamashika2024, rvc24, klobas2024,yamashika2025,ares2025c,yamashika2025b,parez2026,travaglino2026, purvaash26, vescovo26,liu2024,Turkeshi2025,Foligno2025,yu2025b,aditya2025,Bhore2025,yu2025c,yamashika2026, hallam26, li2026,mcroberts2026,yamashika2026b,muller2026,banerjee2025, benini2025,liu2025, aditya26, yu2025, calabrese2026} and open~\cite{Nava2019,Carollo2021,Chatterjee2023,Moroder2024,Nava2024,strachan2025,longhi2025,Nava2025,westhoff2025,beato2026,boubakour2025,caceffo2024,ares2025b,giulio2025,russotto2026,hammer26} settings, and has also been observed experimentally~\cite{Joshi2024,Shapira2024,zhang2025,Xu2026}. Beyond its fundamental interest, potential applications to quantum technologies have begun to be explored, particularly as a route toward accelerating state preparation in quantum simulators~\cite{teza2025, ares2025, westhoff2025,beato2026,boubakour2025}.

In systems with a global conserved charge, the QME can be studied through the entanglement asymmetry~\cite{ares2023, bartlett07, vaccaro08, gour09, marvian14}. This quantum-information observable quantifies the breaking of the symmetry in a subsystem and has been used to characterize the QME and related nonequilibrium phenomena~\cite{ares2023b, hara2026, ferro24, maric25, ferro26,  khor23, ares2024,  mazzoni2026, fossati26, Klobas2025, gibbins2025, ares25circ, yang2026, castroalvaredo26}. For an initially symmetry-breaking state, charge-conserving dynamics preserve the symmetry breaking globally, while a finite subsystem gradually restores the symmetry as it approaches the stationary state. The entanglement asymmetry captures this dynamical symmetry restoration and provides a proxy for relaxation, offering a natural framework to study the QME.

The QME can arise from different microscopic mechanisms depending on the underlying dynamics. In integrable systems, for example, the QME can be microscopically understood in terms of the charge transport properties of quasiparticles~\cite{ares2023,Rylands2024,murciano2024,chalas2024,klobas2024, rvc24,yamashika2024,yamashika2025,ares2025c,yamashika2025b,parez2026,travaglino2026, purvaash26, vescovo26}. In non-integrable systems, by contrast, such a quasiparticle picture is generally unavailable, and different  mechanisms can govern the relaxation~\cite{liu2024,Turkeshi2025,Foligno2025,yu2025b,aditya2025,Bhore2025,yu2025c,yamashika2026,mcroberts2026,yamashika2026b,muller2026,banerjee2025,benini2025,liu2025, aditya26, summer26, hallam26, li2026}.

Clifford circuits provide a tractable setting for studying nonequilibrium quantum dynamics. The Gottesman-Knill theorem allows their efficient classical simulation~\cite{Aaronson2004,anders2006,nest2010,Dehaene2003,nest2004,Hostens2005,bravyi2021,hu2022,gidney2021,garcia2014,vinkhuijzen2023,rall2019,fang2024,brandl2026,garner2025,gottesman1997,gottesman1998}, while they can nevertheless exhibit rapid entanglement growth, mixing, and information scrambling~\cite{znidaric2020,Richter2023,Sierant2023,farshi2023,farshi2022,Zhu2017,webb2016,mitsuhashi2023,zhu2016,gutschow2010,gutschow2009}. This combination makes them a useful setting for identifying microscopic mechanisms of relaxation that can be difficult to isolate in generic quantum dynamics. Clifford circuits are also central to quantum error correction~\cite{gottesman1997,gottesman1998,calderbank1998,dennis2002} and have become an important framework for studying monitored many-body dynamics~\cite{li2018,li2019,choi2020,gullans2020,zabalo2020,lunt2021,sierant2022, russotto26monitored}.

In this Letter, we study the QME in random Clifford circuits with a conserved $\mathrm{U}(1)$ charge. Starting from stabilizer product states, we find that states with stronger initial symmetry breaking can restore the symmetry faster at the subsystem level. We explain this behavior by mapping the Clifford dynamics to a charge-conserving quantum automaton, whose entanglement asymmetry can be effectively described by two particle species undergoing a  symmetric simple exclusion process (SSEP) dynamics. This particle picture yields an analytic expression for the entanglement asymmetry and shows how the initial symmetry breaking sets the particle density, which in turn controls the rate of symmetry restoration.

\textit{Setup}.---
We consider a chain of $N$ qubits evolved by a random Clifford brickwork circuit with a conserved $\mathrm{U}(1)$ charge. We denote the Pauli operators on site $i$ by $X_i$, $Y_i$, and $Z_i$, and choose the computational basis such that $Z_i\ket*{0}_i=\ket*{0}_i$ and $Z_i\ket*{1}_i=-\ket*{1}_i$. Each two-qubit gate $U_{i,i+1}$ is drawn independently from the set of Clifford gates satisfying $[U_{i,i+1},Z_{i}+Z_{i+1}]=0$, which conserves the charge $Q=\sum_{i=1}^N Z_i$. One time step consists of two layers acting on even and odd bonds.

The system is initialized in the following state
\begin{align}\label{eq:Psi_initial}
    \ket*{\Psi(0)}
    =
    \bigotimes_{i\in \Lambda_Z}\ket*{0}_i
    \bigotimes_{j\in \bar{\Lambda}_Z}\ket*{+}_j,
\end{align}
where $\ket*{\pm}_j=(\ket*{0}_j\pm\ket*{1}_j)/\sqrt{2}$, $\Lambda_Z$ is the set of $M$ sites initialized in $\ket*{0}_i$, and $\bar{\Lambda}_Z$ is its complement. The initial state has therefore charge density $m=\ev*{Q/N}{\Psi(0)}=M/N$. For each circuit realization, we independently choose the set $\Lambda_Z$ uniformly at random among all sets of $M$ sites. 
For $m=1$, the initial state reduces to $\bigotimes_i\ket*{0}_i$, which respects the $\mathrm{U}(1)$-rotational symmetry generated by $Q$. As $m$ decreases, the fraction of sites initialized in $\ket*{+}_i$ increases, and the initial state breaks the $\mathrm{U}(1)$ symmetry more strongly. 

We are interested in the fate of the initially broken symmetry during the time evolution. Denoting by $U(t)$ the unitary operator of the full circuit after $t$ time steps, the evolved state is $\ket*{\Psi(t)}=U(t)\ket*{\Psi(0)}$. Since the dynamics is unitary and respects the symmetry, i.e., $[U(t),Q]=0$, a symmetry-breaking initial state cannot evolve into a globally symmetric state. However, the symmetry is expected to be restored locally at long times. We therefore partition the system into $A=\{1,\ldots,N_A\}$ and its complement $\bar A$, and describe $A$ by the reduced density matrix
\begin{align}\label{eq:rho_A_def}
    \rho_A(t)=\Tr_{\bar A}\!\left(\ketbra*{\Psi(t)}\right),
\end{align}
where $\Tr_{\bar A}$ denotes the partial trace over $\bar A$. Under generic circumstances, $\rho_A(t)$ recovers the $\mathrm{U}(1)$ rotational symmetry generated by $Q_A$ at long times~\cite{Polkovnikov-2011,calabrese2016}.

\textit{Subsystem dynamics}.---
Given the initial state~\eqref{eq:Psi_initial} and the Clifford nature of the circuit gates, the dynamics of $\rho_A(t)$ can be described within the stabilizer formalism. If we denote by $\mathcal{S}(t)$ the stabilizer group of $\ket*{\Psi(t)}$, i.e. the group of Pauli strings that leave the state invariant, Eq.~\eqref{eq:rho_A_def} can be rewritten as
\begin{align}\label{eq:rho_A_stabilizer}
    \rho_A(t)
    =
    \frac{1}{2^{N_A}}\sum_{S\in \mathcal{S}_A(t)} S
    =
    \frac{1}{2^{N_A}}\prod_{g\in\mathcal{G}_A(t)}(I+g).
\end{align}
Here, $\mathcal{S}_A(t)\subset \mathcal{S}(t)$ is the subgroup of stabilizers acting trivially on $\bar A$, and $\mathcal{G}_A(t)$ is a generating set for $\mathcal{S}_A(t)$.

\begin{figure}
\raggedleft
\includegraphics[width=0.47\textwidth]{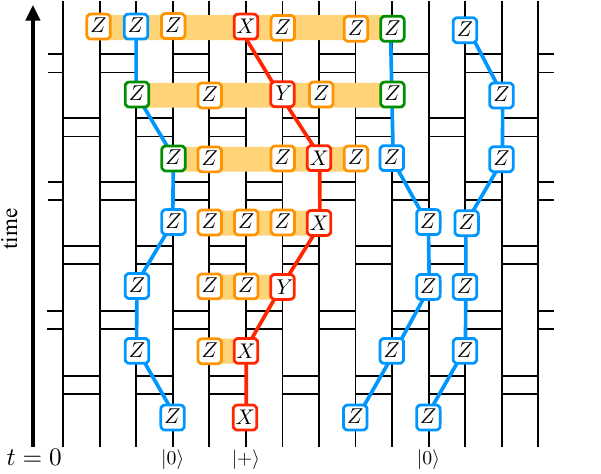}
\caption{Illustration of the stabilizer dynamics under the $\mathrm{U}(1)$-symmetric Clifford brickwork circuit. A $Z_i$ generator associated with a site initialized in $\ket{0}_i$ remains a single-site $Z$ operator, whose position undergoes a random walk. An $X_i$ generator associated with a site initialized in $\ket{+}_i$ retains a single $X$ or $Y$ operator while accumulating $Z$ factors, causing its support to spread. Green $Z$ operators indicate sites where a $Z$ factor of the spreading generator overlaps with a single-site $Z$ stabilizer.}
\label{fig:circuit}
\end{figure}

As shown by Eq.~\eqref{eq:rho_A_stabilizer}, the subsystem dynamics is fully encoded in the time evolution of the stabilizer generators. For the initial state in Eq.~\eqref{eq:Psi_initial}, $\mathcal{G}_A(0)$ reads 
\begin{align}
\mathcal{G}_A(0)
=
\qty{Z_i\,|\,i\in A\cap\Lambda_Z}
\cup
\qty{X_i\,|\,i\in A\cap\bar{\Lambda}_Z}.
\label{eq:initial_subsystem_generators}
\end{align}
The $Z_i$ and $X_i$ operators in $\mathcal{G}_A(0)$ evolve in qualitatively different ways under the $\mathrm{U}(1)$-symmetric Clifford dynamics, as illustrated in Fig.~\ref{fig:circuit}. Each $X_i$ operator evolves into a Pauli string containing a single $X$ or $Y$ operator dressed by $Z$ operators, with its support growing in time.
Once their support extends beyond $A$, the corresponding generators no longer belong to $\mathcal{G}_A(t)$.
In contrast, the $Z_i$ operators remain single-site operators whose position performs a random walk, and thus belong to $\mathcal{G}_A
(t)$ whenever they lie in $A$.  As a result, the reduced density matrix at large times  takes the form 
\begin{align}\label{eq:rho_A_late_time}
    \rho_A(t\gg 1)\simeq \frac{1}{2^{N_A-k}}\prod_{a=1}^k \frac{1+Z_{i_a}}{2}, 
\end{align}
for any initial state and circuit realization, where $k\simeq mN_A$ is the number of $Z_i$ operators inside $A$ and $i_a\in A$ are their positions. 

\begin{figure*}[t]
\includegraphics[width=\textwidth]{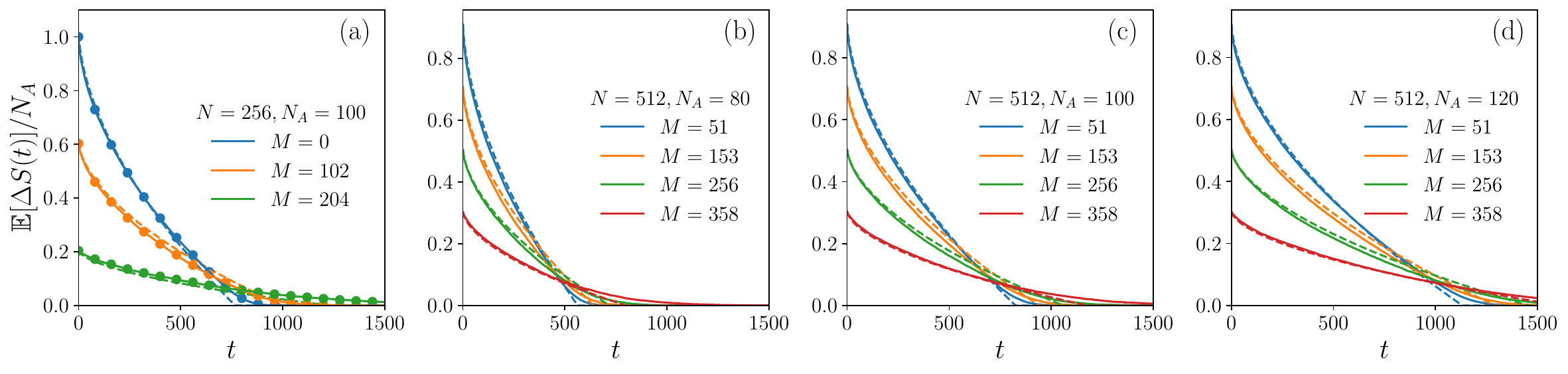}
\caption{Time evolution of the entanglement asymmetry for a subsystem of size $N_A$ in a chain of $N$ qubits. In all panels, solid lines show the numerical results for the random $\mathrm{U}(1)$-symmetric Clifford circuit, while dashed lines show the analytic result in Eq.~\eqref{eq:EA_final}. The symbols in (a) show the numerical results for the quantum automaton defined by Eq.~\eqref{eq:czswap_gate_set}. All numerical data were obtained by averaging over $1000$ choices of $\Lambda_Z$, the set of $M$ sites initialized in the state $\ket{0}$, and circuit realizations.}
\label{fig:EA_dynamics}
\end{figure*}

Equation~\eqref{eq:rho_A_late_time} shows that the subsystem restores at long times not only the global $\mathrm{U}(1)$ symmetry generated by $Q_A=\sum_{i\in A} Z_i$, but also the local $\mathrm{U}(1)$ symmetry generated by $\{Z_i\}_{i\in A}$, even though the dynamics conserves only the global charge. The emergence of this larger symmetry is a consequence of the restricted structure of stabilizer states, whose reduced density matrices have flat spectra and therefore cannot reproduce the Gibbs state $\rho_A \propto \Tr_{\bar A} e^{-\lambda Q}$ expected at long times in generic global charge-conserving dynamics.

\textit{Entanglement asymmetry}.---
We quantify the restoration of the local $\mathrm{U}(1)$ symmetry using the entanglement asymmetry~\cite{ares2023}
\begin{align}
\Delta S(t)
=
S(\Pi[\rho_A(t)])-S(\rho_A(t)),
\label{eq:EA_definition}
\end{align}
where $S(\rho)=-\Tr(\rho\log_2\rho)$ is the von Neumann entropy and $\Pi[\bullet]=\sum_{\mathbf q}\Pi_{\mathbf q} \bullet \Pi_{\mathbf q}$ denotes the symmetrization with respect to the local $\mathrm{U}(1)$ symmetry. Here $\Pi_{\mathbf q}=\prod_{i\in A}(I+q_iZ_i)/2$ is the projector onto the eigenspace of $Z_i$ with eigenvalue $q_i=\pm1$. The entanglement asymmetry faithfully measures the breaking of the local $\mathrm{U}(1)$ symmetry at the subsystem level, satisfying $\Delta S(t) \geq 0$, with $\Delta S(t) = 0$ if and only if $[\rho_A(t),Z_i]=0$ for every $i\in A$~\cite{ma2022,han2023}.

In our case, according to Eq.~\eqref{eq:rho_A_late_time}, the entanglement asymmetry vanishes at long times for any circuit realization and initial state. We therefore investigate how the relaxation of the averaged entanglement asymmetry depends on the initial charge density $m$ and whether this leads to a Mpemba effect. The QME is characterized by an inversion of the relaxation ordering: a state with stronger initial symmetry breaking, $\mathbb{E}[\Delta S_\one(0)]>\mathbb{E}[\Delta S_\two(0)]$, 
eventually becomes less asymmetric, $\mathbb{E}[\Delta S_\one(t)]<\mathbb{E}[\Delta S_\two(t)]$ for $t>t_{\rm M}$~\cite{ares2023}.

\textit{Entanglement asymmetry of the stabilizer states}.---
Using the standard formula $S=N_A-|\mathcal{G}_A|$ for the entanglement entropy of a stabilizer state~\cite{fattal2004}, the averaged entanglement asymmetry can be written in terms of the stabilizer generators as 
\begin{align}\label{eq:EA_stabilizer}
    \mathbb{E}[\Delta S_A(t)]=\mathbb{E}[S(\Pi[\rho_A(t)])]-N_A +\mathbb{E}[|\mathcal{G}_A(t)|]. 
\end{align}
The first term in this equation can be directly obtained from Eq.~\eqref{eq:rho_A_stabilizer}. The symmetrization $\Pi[\bullet]$ eliminates all Pauli strings that are not diagonal in the local $Z$ basis. Hence, all stabilizers containing $X$ or $Y$ operators are removed from Eq.~\eqref{eq:rho_A_stabilizer}. Since the generators evolved from the $X_i$ operators in the initial state~\eqref{eq:initial_subsystem_generators} necessarily contain an $X$ or $Y$ component, only the single-site $Z_i$ generators contribute to $\Pi[\rho_A(t)]$, leading to
\begin{align}
\Pi[\rho_A(t)]
=
\frac{1}{2^{N_A}}
\prod_{i\in A\cap\Lambda_Z(t)}(I+Z_i), 
\label{eq:symmetrized_density_matrix}
\end{align}
where $\Lambda_Z(t)$ denotes the set of sites occupied by single-site $Z$ stabilizers at time $t$. The entropy of the state in Eq.~\eqref{eq:symmetrized_density_matrix} is therefore $S(\Pi[\rho_A(t)])=N_A-|A\cap\Lambda_Z(t)|$ and, since each site belongs to $\Lambda_Z(t)$ with probability $m$, the averaged entropy is $\mathbb{E}[S(\Pi[\rho_A(t)])]=N_A(1-m)$, which is time-independent.
Plugging this result into Eq.~\eqref{eq:EA_stabilizer}, we obtain 
\begin{align}\label{eq:EA_relation}
    \mathbb{E}[\Delta S(t)]
    =
    \mathbb{E}[|\mathcal{G}_A(t)|]-N_A m.
\end{align}
Thus, the time evolution of the entanglement asymmetry is uniquely determined by $|\mathcal{G}_A(t)|$.
Numerically, $|\mathcal{G}_A(t)|$ can be efficiently computed by identifying the independent stabilizers whose support is entirely contained in $A$~\cite{Aaronson2004}.

Figure~\ref{fig:EA_dynamics} shows the time evolution of the averaged entanglement asymmetry for several charge densities. The curves exhibit an inversion of the relaxation ordering: a state with stronger initial symmetry breaking (smaller $m$) becomes less asymmetric at late times than a more symmetric initial state. This demonstrates the occurrence of the Mpemba effect in the $\mathrm{U}(1)$-symmetric Clifford circuit for the family of states~\eqref{eq:Psi_initial}.

\textit{Reduction to a quantum automaton}.---
To elucidate the origin of the QME observed in Fig.~\ref{fig:EA_dynamics}, we analytically determine $|\mathcal{G}_A(t)|$, the only time-dependent quantity entering Eq.~\eqref{eq:EA_relation}. This is equivalent to counting the number of stabilizers that act trivially on $\bar A$, with $|\mathcal{S}_A(t)|=2^{|\mathcal{G}_A(t)|}$. When constructing elements of the restricted stabilizer group $\mathcal{S}_A(t)$ from products of generators, two conditions must be satisfied. 
All non-$Z$ components must lie within $A$, while any $Z$ operator in $\bar{A}$ must be canceled by another $Z$ operator. These conditions depend 
only on the positions of the single-site $Z$ generators and on the support of the spreading generators, but not on whether the non-$Z$ component of 
each generator is $X$ or $Y$.

Exploiting this observation, we can describe the time evolution of the entanglement asymmetry in the original circuit by a simpler brickwork circuit consisting of the two-site gate
\begin{align}
V_{i,i+1}
\in
\{I,\mathrm{CZ},\mathrm{SWAP},\mathrm{SWAP}\,\mathrm{CZ}\},
\label{eq:czswap_gate_set}
\end{align}
where $\mathrm{CZ}$ is the controlled-$Z$ gate and $\mathrm{SWAP}$ exchanges the two qubits. The four gates are chosen with probability $1/4$. 
The effective circuit preserves the global $\mathrm{U}(1)$ charge $Q$ and reproduces the stochastic evolution of the single-site $Z$ generators and the support of the spreading generators~\cite{SM}. Since this information fixes $|\mathcal{G}_A(t)|$, the original and effective circuits generate the same statistics of $|\mathcal{G}_A(t)|$, and therefore the same averaged entanglement entropy and asymmetry, as confirmed numerically in Fig.~\ref{fig:EA_dynamics}(a).

The advantage of the effective circuit is that it is a quantum automaton~\cite{gopalakrishnan2018,Iaconis2019}, which maps each computational-basis state to another computational-basis state up to a phase. Specifically, denoting its time evolution operator by $V(t)$, the corresponding time-evolved state can be written as
\begin{align}
V(t)\ket{\Psi(0)}
=
\frac{1}{\sqrt{|\mathcal{C}|}}
\sum_{a\in\mathcal{C}}
e^{\im \phi_t(a)}
\ket{a(t)},
\label{eq:qa_wavefunction}
\end{align}
where $\mathcal{C}$ is the set of bit strings appearing in the initial state. Each $a\in\mathcal{C}$ is written as $a=a_1a_2\cdots a_N$ with $a_i\in\{0,1\}$. We denote by $a(t)$ the bit string obtained from $a$ after applying the SWAP gates throughout the evolution, while $\phi_t(a)$ denotes the phase accumulated from the CZ gates.

\begin{figure}
\includegraphics[width=0.48\textwidth]{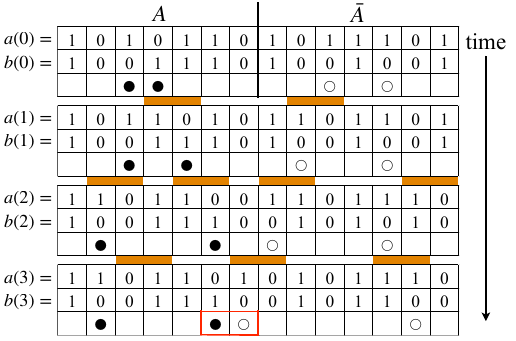}
\caption{Schematic illustration of the two-species particle model. $a(t)$ and $b(t)$ are two possible bit strings in the state~\eqref{eq:qa_wavefunction} of the quantum automaton at time $t$. Sites in $A$ ($\bar A$) where $a_i(t)\neq b_i(t)$ correspond to black (white) particles. The orange box marks a pair of neighboring sites on which a SWAP gate is applied, exchanging their bit-string pairs, and thereby moving the particles. The red box marks the first adjacency between a black and a white particle, after which the bit-string pair is no longer counted in $\mathcal{F}(t)$.}
\label{fig:two_species}
\end{figure}

For the state~\eqref{eq:qa_wavefunction}, the subsystem purity provides a direct route to determining $|\mathcal{G}_A(t)|$ via $\Tr_A[\rho_A(t)^2]=2^{|\mathcal{G}_A(t)|-N_A}$. We can evaluate this purity using the two-species particle model proposed in Refs.~\cite{Han2022,Han2023Jan,Han2023Feb} for the quantum automaton~\eqref{eq:qa_wavefunction}. The model considers pairs of bit strings $(a,b)\in\mathcal{C}^2$. For each pair, a black particle is placed at every site $i\in A$ where $a_i\neq b_i$, while a white particle is placed at every site $i\in \bar{A}$ where $a_i\neq b_i$, as illustrated in Fig.~\ref{fig:two_species}. 
As the SWAP gates in $V(t)$ update $a(t)$ and $b(t)$, the black/white particle positions evolve accordingly, as shown in the same figure. 

As explicitly shown in SM~\cite{SM}, see also Refs.~\cite{Han2022,Han2023Jan,Han2023Feb}, the two-species particle model predicts $\Tr_A[\rho_A(t)^2]\simeq \mathcal{F}(t)$, where $\mathcal{F}(t)$ is the fraction of bit-string pairs $(a,b)\in \mathcal{C}^2$ for which black and white particles have never become adjacent up to time $t$.
Using this result together with Eq.~\eqref{eq:EA_relation}, we obtain
\begin{align}
\mathbb{E}[\Delta S(t)]
&\simeq 
N_A(1-m)+\mathbb{E}\qty[\log_2 \mathcal{F}(t)].
\label{eq:EA_purity}
\end{align}
Equation~\eqref{eq:EA_purity} shows that the decay of the entanglement asymmetry, and hence the restoration of the symmetry, is determined by the two-species particle dynamics encoded in $\mathcal{F}(t)$. As time progresses, particles move so that, for an increasing number of bit-string pairs, a black particle becomes adjacent to a white particle. These bit-string pairs are then excluded from $\mathcal{F}(t)$. The resulting decrease of $\mathcal{F}(t)$ makes the logarithmic term in Eq.~\eqref{eq:EA_purity} smaller and reduces the entanglement asymmetry.

\textit{Entanglement asymmetry from SSEP}.---
The fraction $\mathcal{F}(t)$ can be evaluated from the particle density in the two-species model and their motion. When a bit-string pair $(a,b)$ is randomly drawn from $\mathcal{C}^2$, the bits $a_i$ and $b_i$ are independently and uniformly chosen on each site $i\in\bar{\Lambda}_Z$. Thus, on each of the $N-M$ sites in $\bar{\Lambda}_Z$ , the probability that $a_i \neq b_i$ is $1/2$. The particle density is therefore $n(m)=(N-M)/(2N)=(1-m)/2$. The fraction of bit-string pairs with $r$ empty sites between the nearest black and white particles is $n(m)^2[1-n(m)]^r$. 
Let $r_*(t)$ denote the characteristic initial number of empty sites separating the nearest black and white particles below which they typically become adjacent by time $t$. Thus, pairs with $r<r_*(t)$ have typically become adjacent by time $t$, whereas those with $r>r_*(t)$ have typically not.
Thus, pairs with $r<r_*(t)$ typically have become adjacent by time $t$, whereas pairs with $r>r_* (t)$ typically have not yet become adjacent. The pairs counted in $\mathcal{F}(t)$ therefore satisfy $r\geq r_*(t)$, yielding
\begin{align}
\mathbb{E}[\mathcal{F}(t)]
\simeq
\sum_{r\geq r_*(t)}
n(m)^2[1-n(m)]^r.
\label{eq:no_contact_sum}
\end{align}
We can evaluate $r_*(t)$ using the fact that the particles obey the SSEP generated by the SWAP gates in $V(t)$. In particular, we find $r_*(t)=\sqrt{2t\ln t}$ at the leading order in $t$~\cite{SM}.
Applying this result in Eq.~\eqref{eq:no_contact_sum} with the assumption $\mathbb{E}[\log_2 \mathcal{F}(t)]\simeq \log_2\mathbb{E}[\mathcal{F}(t)]$, we finally obtain 
\begin{multline}\label{eq:EA_final}
\mathbb{E}[\Delta S(t)]
\simeq
\max\Bigl[
0,\,
N_A(1-m)
\\
+\sqrt{2t\ln t}\,\log_2\qty[1-n(m)]
\Bigr],
\end{multline}
where we introduced the maximum function to enforce the non-negativity of the entanglement asymmetry. 
This result gives the leading-order behavior of the entanglement asymmetry in the large-$t$ and large-$N_A$ limit.

The dashed curves in Fig.~\ref{fig:EA_dynamics} show the analytic prediction of Eq.~\eqref{eq:EA_final}. They agree well with the exact numerical results for the random U(1)-symmetric Clifford circuit (solid curves) and reproduce the inversion of the relaxation ordering, in which initially more asymmetric states eventually have smaller entanglement asymmetry than initially less asymmetric states. This agreement confirms that the two-species particle model captures the dynamics of the entanglement asymmetry and the QME.

\textit{Origin of the QME.---}
Equation~\eqref{eq:EA_final} explains the QME in terms of the two-species particle model. Consider two charge densities $m_\one<m_\two$. On the one hand, since the ensemble of initial states with $m_\one$ has more sites initialized in $\ket*{+}$, its initial entanglement asymmetry is larger, $\mathbb{E}[\Delta S_\one(0)]>\mathbb{E}[\Delta S_\two(0)]$. On the other hand, the same ensemble has a larger particle density, $n(m_\one)>n(m_\two)$. The larger particle density makes the typical distance between black and white particles shorter, increasing the number of bit-string pairs in which a black and a white particle have become adjacent by time $t$. As a result, $\mathcal{F}(t)$ decreases faster for $m_\one$, and Eq.~\eqref{eq:EA_purity} gives a faster decay of the entanglement asymmetry. Consequently, there is a time $t_\mathrm{M}$ after which $\mathbb{E}[\Delta S_\one(t)]<\mathbb{E}[\Delta S_\two(t)]$. 

We finally note that the analytic result in Eq.~\eqref{eq:EA_final} is not specific to the initial state in Eq.~\eqref{eq:Psi_initial}. In~\cite{SM}, we show that Eq.~\eqref{eq:EA_final} also holds for a general product stabilizer initial state. In this case, $\Lambda_Z$ denotes the set of sites initialized in either $\ket{0}_i$ or $\ket{1}_i$ , while the remaining sites are initialized in one of the $X$- or $Y$-eigenstates. The parameter $m$ is now the fraction of sites belonging to $\Lambda_Z$. The evolution of the stabilizer supports depends only on which sites are initialized in computational-basis states, and not on the specific choice of stabilizer state on the remaining sites. The resulting entanglement asymmetry therefore remains unchanged, and the same mechanism leads to the QME.

\textit{Conclusions}.---
In this Letter, we investigated the QME in U(1)-symmetric random Clifford circuits, showing that the entanglement asymmetry can be effectively described by two particle species undergoing SSEP dynamics. In this picture, symmetry restoration is controlled by how frequently particles of different species encounter each other. This particle description yields an analytic expression for the entanglement asymmetry and explains the QME: a denser particle configuration corresponds to larger initial symmetry breaking, but also makes encounters between particles of different species more likely, leading to faster symmetry restoration. Our results reveal a distinct mechanism for the QME and provide an intuitive and quantitative picture of symmetry relaxation in Clifford dynamics through the underlying SSEP.

An important direction is to determine how robust and general this picture is beyond the unitary Clifford circuits studied here. Measurements could modify the particle dynamics and thereby alter the symmetry restoration. More generally, studying symmetric quantum automata that are not Clifford circuits could reveal whether this particle picture survives in the absence of stabilizer structure.
\\~\par
S.Y. acknowledges support from JSPS KAKENHI Grants No.~JP25K23355 and JP26K17050, and from the Institute for Advanced Science, University of Electro-Communications.   P.C.  and F.A. acknowledge support from the European Research Council
under the Advanced Grant no.~101199196 (MOSE).

\bibliography{ref} 

\newpage

\usetikzlibrary{decorations.pathreplacing}

\setcounter{page}{1}
\setcounter{secnumdepth}{2}
\onecolumngrid
\newcounter{equationSM}
\newcounter{figureSM}
\newcounter{tableSM}
\stepcounter{equationSM}
\setcounter{equation}{0}
\setcounter{figure}{0}
\setcounter{table}{0}
\setcounter{section}{0}
\makeatletter
\renewcommand{\theequation}{\textsc{sm}-\arabic{equation}}
\renewcommand{\thefigure}{\textsc{sm}-\arabic{figure}}
\renewcommand{\thetable}{\textsc{sm}-\arabic{table}}
\renewcommand{\thesection}{\Roman{section}}
\makeatother

\begin{center}
{\Large\bf Supplemental Material for\\
``The quantum Mpemba effect in symmetric random Clifford circuits''}
\end{center} 

\noindent
In this supplemental material, we report some useful information complementing the main text:
\begin{itemize}
\item In Sec.~\ref{sec:sm_czswap_reduction}, we describe the original random $\mathrm{U}(1)$-symmetric Clifford circuit and explain its reduction to the quantum automaton.
\item In Sec.~\ref{sec:sm_purity_phase}, we derive the phase-sum representation of the purity.
\item In Sec.~\ref{sec:sm_two_species}, we rewrite the phase sum as a two-species particle model.
\item In Sec.~\ref{sec:sm_leading_particle}, we estimate the probability that the two particle species do not become adjacent using exclusion-process statistics.
\item In Sec.~\ref{sec:sm_general_initial_states}, we discuss the generality of the quantum Mpemba effect for product stabilizer initial states.
\end{itemize}

\section{Mapping to the quantum automaton}
\label{sec:sm_czswap_reduction}

In this section, we explain the reduction of the random $\mathrm{U}(1)$-symmetric Clifford circuit considered in the main text to the quantum automaton. In the random $\mathrm{U}(1)$-symmetric Clifford brickwork circuit, one time step consists of two layers acting on even and odd bonds, and each local gate is drawn independently from the two-qubit Clifford gates that commute with $Z_i+Z_{i+1}$. Consequently, the total charge $Q=\sum_{i=1}^N Z_i$ is conserved. A convenient parametrization of this gate set is~\cite{Richter2023}
\begin{align}
U_{i,i+1}
=
\begin{cases}
(A_i \otimes A_{i+1})(C_i\otimes C_{i+1}),\\
(A_i \otimes B_{i+1})\mathrm{CX}_{i,i+1}(I_i\otimes W_{i+1})(C_i\otimes C_{i+1}),\\
(B_i\otimes A_{i+1})\mathrm{CX}_{i,i+1}\mathrm{CX}_{i+1,i}(I_i\otimes W_{i+1})(C_i\otimes C_{i+1}),\\
(A_i\otimes A_{i+1})\mathrm{CX}_{i,i+1}\mathrm{CX}_{i+1,i}\mathrm{CX}_{i,i+1}(C_i\otimes C_{i+1}),
\end{cases}
\label{eq:sm_original_clifford_gate}
\end{align}
where $A\in\{I,P\}$, $B\in\{V,H\}$, and $C\in\{I,Z\}$. Here,
\begin{align}
P=\begin{pmatrix}1&0\\0&\im\end{pmatrix},\quad
H=\frac{1}{\sqrt{2}}\begin{pmatrix}1&1\\1&-1\end{pmatrix},
\quad
W=HP,\quad
V=W^2.
\label{eq:sm_single_qubit_gates}
\end{align}
The controlled-NOT gates are
\begin{align}
\mathrm{CX}_{i,i+1}
=
\begin{pmatrix}
1&0&0&0\\
0&1&0&0\\
0&0&0&1\\
0&0&1&0
\end{pmatrix},
\qquad
\mathrm{CX}_{i+1,i}
=
\begin{pmatrix}
1&0&0&0\\
0&0&0&1\\
0&0&1&0\\
0&1&0&0
\end{pmatrix}.
\label{eq:sm_cx_gates}
\end{align}
In the main text, the initial state of the circuit is
\begin{align}\label{eq:ini_st_sm}
    \ket*{\Psi(0)}
    =
    \bigotimes_{i\in \Lambda_Z}\ket*{0}_i
    \bigotimes_{j\in \bar{\Lambda}_Z}\ket*{+}_j,
\end{align}
where $\ket*{\pm}_j=(\ket*{0}_j\pm\ket*{1}_j)/\sqrt{2}$, $\Lambda_Z$ is the set of $M$ sites initialized in $\ket*{0}_i$, and $\bar{\Lambda}_Z$ is its complement. The $Z$-type stabilizer generators, associated with the sites initially in the state $\ket*{0}_i$, remain local under this circuit.  
Specifically, under the update of a bond in a given brickwork layer, they evolve as
\begin{align}
\left.
\begin{array}{ll}
Z\otimes I\\
I\otimes Z
\end{array}
\right\}
\mapsto
\left\{
\begin{array}{ll}
Z\otimes I &(\text{prob. }1/2)\\
I\otimes Z &(\text{prob. }1/2)
\end{array}
\right.,
\qquad
Z\otimes Z\mapsto Z\otimes Z.
\label{eq:sm_Z_update}
\end{align}
Thus the local $Z$ generators perform random walks with exclusion. We denote as $\Lambda_Z(t)$ the set of sites where these $Z$ generators are supported at time $t$. On the other hand, an $X$-type stabilizer generator, associated with a site prepared in the state $\ket{+}_i$, remains a Pauli string with exactly one $X$ or $Y$ operator, and with only $I$ or $Z$ operators on all other sites. Specifically, for $P,P'\in\{X,Y\}$, its transformation under a local gate in a given brickwork layer is
\begin{align}
\left.
\begin{array}{ll}
P\otimes I\\
I\otimes P\\
P\otimes Z\\
Z\otimes P
\end{array}
\right\}
\mapsto
\left\{
\begin{array}{ll}
P'\otimes I &(\text{prob. }1/4)\\
I\otimes P' &(\text{prob. }1/4)\\
P'\otimes Z &(\text{prob. }1/4)\\
Z\otimes P' &(\text{prob. }1/4)
\end{array}
\right. .
\label{eq:sm_P_update}
\end{align}
More explicitly, these generators can be written in the form
\begin{align}
g(t)
=
P_{x(t)}
\prod_{j\in\Gamma(t)}Z_j,
\qquad
P_{x(t)}\in\{X_{x(t)},Y_{x(t)}\},
\label{eq:sm_pauli_string_structure}
\end{align}
where $x(t)$ is the site carrying the unique $X$ or $Y$ operator and $\Gamma(t)$ is the set of sites where the string contains a $Z$ operator. Multiplying $g(t)$ by a single-site $Z$ generator at a site $j\in\Lambda_Z(t)$ changes whether $Z_j$ is present in the product in Eq.~\eqref{eq:sm_pauli_string_structure}. Therefore, the question of whether a product of stabilizers can be supported entirely in $A$ depends on $x(t)$ and on $\Gamma(t)$ modulo the single-site $Z$ generators of the set $\Lambda_Z(t)$. It does not depend on whether the operator at $x(t)$ is $X$ or $Y$, nor on the overall sign of the stabilizer. To calculate $|\mathcal{G}_A(t)|$, we may therefore represent the operator at $x(t)$ by $X$:
\begin{align}
\left.
\begin{array}{ll}
X\otimes I\\
I\otimes X\\
X\otimes Z\\
Z\otimes X
\end{array}
\right\}
\mapsto
\left\{
\begin{array}{ll}
X\otimes I &(\text{prob. }1/4)\\
I\otimes X &(\text{prob. }1/4)\\
X\otimes Z &(\text{prob. }1/4)\\
Z\otimes X &(\text{prob. }1/4)
\end{array}
\right. .
\label{eq:sm_effective_update}
\end{align}
This effective update is generated by the random two-qubit Clifford gate in the quantum automaton,
\begin{align}
V_{i,i+1}
=
\begin{cases}
I &(\text{prob. }1/4),\\
\mathrm{CZ}_{i,i+1} &(\text{prob. }1/4),\\
\mathrm{SWAP}_{i,i+1} &(\text{prob. }1/4),\\
\mathrm{SWAP}_{i,i+1}\mathrm{CZ}_{i,i+1} &(\text{prob. }1/4).
\end{cases}
\label{eq:sm_czswap_gate}
\end{align}
The equivalence follows directly from the local conjugation rules. The CZ gate leaves all $Z$ operators invariant and attaches a $Z$ to each neighboring $X$,
\begin{align}
\mathrm{CZ}^\dagger(X\otimes I)\mathrm{CZ}
=
X\otimes Z,
\qquad
\mathrm{CZ}^\dagger(I\otimes X)\mathrm{CZ}
=
Z\otimes X,
\label{eq:sm_cz_conjugation}
\end{align}
while SWAP exchanges the two sites. Hence the four equally likely gates in Eq.~\eqref{eq:sm_czswap_gate} give
\begin{align}
\begin{array}{c|c|c}
V_{i,i+1} & Z\otimes I\mapsto & X\otimes I\mapsto\\
\hline
I & Z\otimes I & X\otimes I\\
\mathrm{CZ} & Z\otimes I & X\otimes Z\\
\mathrm{SWAP} & I\otimes Z & I\otimes X\\
\mathrm{SWAP}\,\mathrm{CZ} & I\otimes Z & Z\otimes X
\end{array}.
\label{eq:sm_czswap_conjugation_table}
\end{align}
The table for $I\otimes Z$ and $I\otimes X$ follows by exchanging left and right. Therefore, the quantum automaton reproduces Eqs.~\eqref{eq:sm_Z_update} and \eqref{eq:sm_effective_update}. It also preserves the exclusion of local $Z$ walkers, because $Z\otimes Z$ is invariant under all four gates. Thus the quantum automaton reproduces the stabilizer-support dynamics relevant for determining $|\mathcal{G}_A(t)|$ in the original $\mathrm{U}(1)$-symmetric Clifford circuit, although it is not a microscopic replacement of every amplitude in the original circuit.

The computational-basis dynamics of the quantum automaton is classical, with phases attached to the basis states. Associating each computational-basis state with an $N$-bit string $a=a_1\cdots a_N$, where $a_i\in \{0,1\}$, the gates  $\mathrm{SWAP}$ only permute the bits, whereas a $\mathrm{CZ}$ gate acting on sites $i$ and $i+1$ multiplies the state by the phase factor $(-1)^{a_i a_{i+1}}$. We denote by $a(t)$ the bit string obtained from $a$ after the SWAP gates up to time $t$, and by $\phi_t(a)$ the phase accumulated from the $\mathrm{CZ}$ gates during this evolution. Then the time evolution of the computational-basis states in the quantum automaton is
\begin{align}
V(t)\ket*{a}
=
e^{\im\phi_t(a)}\ket*{a(t)},
\label{eq:sm_phase_function}
\end{align}
where $\phi_t(a)$ is $\pi$ times the number of $\mathrm{CZ}$ gates that acted on pairs of occupied bits during the evolution of the initial string $a$, modulo $2\pi$. Therefore, $\phi_t(a)=0, \pi$.

\section{Entanglement entropy as a phase sum}
\label{sec:sm_purity_phase}

Using the map to the quantum automaton obtained in Sec.~\ref{sec:sm_czswap_reduction}, we now derive an expression for the entanglement entropy of the original random $\mathrm{U}(1)$ Clifford circuit in terms of the phases~\eqref{eq:sm_phase_function} accumulated during the evolution in the corresponding quantum automaton.

Let $\mathcal{C}$ be the set of bit strings appearing in the initial product state~\eqref{eq:ini_st_sm}. 
Since $M$ sites are initialized in the state $\ket*{0}$ and the remaining $N-M$ sites in $\ket*{+}$, the bit strings belonging to $\mathcal{C}$ are of the form
\begin{align}
    \mathcal{C}=\{a=a_1a_2...a_N\in \{0,1\}^N\mid a_i=0\quad \forall i \in \Lambda_Z\}. 
\end{align}
The number of bit strings in $\mathcal{C}$ is $|\mathcal{C}|=2^{N-M}$. We denote the time-evolved state of the quantum automaton by $\ket*{\tilde{\Psi}(t)}$. Using Eq.~\eqref{eq:sm_phase_function}, this state can be written as
\begin{align}
\ket*{\tilde{\Psi}(t)}
=
\frac{1}{\sqrt{|\mathcal{C}|}}
\sum_{a\in\mathcal{C}}e^{\im\phi_t(a)}\ket*{a(t)}.
\label{eq:sm_state_basis_map}
\end{align}
Here, $\phi_t(a)=0, \pi$ is the phase accumulated by the CZ gates present in the time evolution operator $V(t)$. Since $V(t)$ consists only of the CZ gates and the SWAP gates, $a(t)$ is related to the initial bit string $a$ by a bit permutation $\pi_t$, such that $a(t)=\pi_t(a)$. Therefore, if we introduce the set of bit strings 
\begin{align}\label{eq:C_t}
    \mathcal{C}_t = \{\pi_t(a)\mid\forall a \in \mathcal{C}\}=\{a=a_1a_2...a_N \in \{0,1\}^N\mid a_i=0\quad \forall i\in \Lambda_Z(t)\}, 
\end{align}
then Eq.~\eqref{eq:sm_state_basis_map} can be rewritten as 
\begin{align}\label{eq:sm_state_basis_map_2}
    \ket*{\tilde{\Psi}(t)}= 
    \frac{1}{\sqrt{\cal |C|}} \sum_{a\in \mathcal{C}_t} 
    e^{\im \phi_t (\pi_t^{-1}(a))}\ket{a}. 
\end{align}

Since the entanglement spectrum of a stabilizer state is flat, the von Neumann entropy equals the second Rényi entropy and can be written in terms of the purity $P(t)=\tr(\rho_A^2)$ as
\begin{align}
S(\rho_A(t))
=
-\log_2 P(t).
\label{eq:sm_entropy_purity}
\end{align}
The purity is obtained from the expectation value of the replica swap operator as
\begin{align}
P(t)
=
\bra*{\tilde{\Psi}(t)}\bra*{\tilde{\Psi}(t)}
\mathrm{SWAP}_A
\ket*{\tilde{\Psi}(t)}\ket*{\tilde{\Psi}(t)}.
\label{eq:sm_swap_purity}
\end{align}
 If $a_A,a_{\bar A}$ and $b_A,b_{\bar A}$ are the restrictions of two bit strings $a$, $b$ to $A$ and $\bar A$, respectively, then the replica swap operator acts as
 \begin{align}
\mathrm{SWAP}_A\ket*{a_Aa_{\bar A}}\ket*{b_Ab_{\bar A}}=
 \ket*{b_Aa_{\bar A}}\ket*{a_Ab_{\bar A}}.
 \end{align}
Substituting Eq.~\eqref{eq:sm_state_basis_map_2} into Eq.~\eqref{eq:sm_swap_purity}, we obtain 
\begin{align}
P(t)
&=
\frac{1}{|\mathcal{C}|^2}
\sum_{a,b,a',b'\in\mathcal{C}_t}
e^{\im\phi_t(\pi_t^{-1}(a))+\im\phi_t(\pi_t^{-1}(b))
-\im\phi_t(\pi_t^{-1}(a'))-\im\phi_t(\pi_t^{-1}(b'))}
\bra*{a'}\bra*{b'}\mathrm{SWAP}_A\ket*{a}\ket*{b}.
\label{eq:sm_purity_complete_basis}
\end{align}
Inserting the two-replica completeness relation, it expands as 
\begin{align}
P(t)=\frac{1}{|\mathcal{C}|^2}
\sum_{c,d\in \{0,1\}^{N}}
\sum_{a,b,a',b'\in\mathcal{C}_t}
e^{\im\phi_t(\pi_t^{-1}(a)+\im\phi_t(\pi_t^{-1}(b))
-\im\phi_t(\pi_t^{-1}(a')-\im\phi_t(\pi_t^{-1}(b'))}
\bra*{a'}\bra*{b'}\mathrm{SWAP}_A\ket{c}\ket{d}\bra{c}\bra{d}\,\ket*{a}\ket*{b}.
\end{align}
If we introduce the swapped bit strings
\begin{align}
\ket*{\tilde c}\ket*{\tilde d}
=
\mathrm{SWAP}_A\ket*{c}\ket*{d},
\qquad
\tilde c=d_A c_{\bar A},
\quad
\tilde d =c_A d_{\bar A},
\label{eq:sm_initial_swapped_strings}
\end{align}
then it reads
\begin{align}
P(t)= 
\frac{1}{|\mathcal{C}|^2}
\sum_{c,d\in \{0,1\}^{N}}
\sum_{a,b,a',b'\in\mathcal{C}_t}
e^{\im\phi_t(\pi_t^{-1}(a))+\im\phi_t(\pi_t^{-1}(b))
-\im\phi_t(\pi_t^{-1}(a'))-\im\phi_t(\pi_t^{-1}(b'))}
\delta_{a',\tilde{c}}\delta_{b',\tilde{d}}\delta_{c,a}\delta_{d,b}. 
\label{eq:sm_purity_complete_basis_expand}
\end{align}
After summing over $a$ and $b$, Eq.~\eqref{eq:sm_purity_complete_basis_expand} reduces to 
\begin{align}
    P(t) 
    = 
    \frac{1}{|\mathcal{C}|^2}
    \sum_{a',b',c,d\in\mathcal{C}_t}
    e^{\im\phi_t(\pi_t^{-1}(c))+\im\phi_t(\pi_t^{-1}(d))
    -\im\phi_t(\pi_t^{-1}(a'))-\im\phi_t(\pi_t^{-1}(b'))}
    \delta_{a',\tilde{c}}\delta_{b',\tilde{d}}.
\label{eq:sm_purity_complete_basis_expand_2}
\end{align}
To perform the sums over $a'$ and $b'$ in Eq.~\eqref{eq:sm_purity_complete_basis_expand_2}, we use the fact that, if the bit strings $c$ and $d$ belong to the set $\mathcal{C}_t$, then the swapped bit strings $\tilde c$ and $\tilde d$ also belong to that set, 
\begin{align}\label{eq:cd}
    c,d\in \mathcal{C}_t \Rightarrow \tilde{c},\tilde{d}\in \mathcal{C}_t. 
\end{align}
This can be shown as follows. When $c,d\in \mathcal{C}_t$, we have that the bits corresponding to the sites $i \in \bar{A}\cap \Lambda_Z(t)$ satisfy 
\begin{align}
    \tilde{c}_i = c_i = 0,\quad \tilde{d}_i = d_i = 0. 
\end{align}
On the other hand, for $i\in A\cap \Lambda_z(t)$, we have
\begin{align}
    \tilde{c}_i=d_i=0,\quad \tilde{d}_i=c_i=0. 
\end{align}
We thus obtain 
\begin{align}
    \tilde{c}_i=\tilde{d}_i = 0\quad \forall i\in \Lambda_Z(t), 
\end{align}
which is the condition that a bit string must satisfy to belong to $\mathcal{C}_t$, according to  Eq.~\eqref{eq:C_t}. 
This result directly implies Eq.~\eqref{eq:cd}. We can then perform the sums over $a'$
and $b'$ in Eq.~\eqref{eq:sm_purity_complete_basis_expand_2}, obtaining
\begin{align}
    P(t)
    &=\frac{1}{|\mathcal{C}|^2}
    \sum_{c,d\in \mathcal{C}_t}
    e^{\im\phi_t(\pi_t^{-1}(c))+\im\phi_t(\pi_t^{-1}(d))
    -\im\phi_t(\pi_t^{-1}(\tilde{c}))-\im\phi_t(\pi_t^{-1}(\tilde{d}))}.
\end{align}
Finally, rewriting $(c, d)$ as $(a, b)$ and using the identity 
\begin{align}
    \sum_{a\in \mathcal{C}_t} f(a)=\sum_{a\in \mathcal{C}} f(\pi_t(a)), 
\end{align}
valid for any function $f$,
we find
\begin{align}
   P(t) =
    \frac{1}{|\mathcal{C}|^2}
    \sum_{a,b\in \mathcal{C}} 
    e^{\im \Phi_t(a,b)},  \label{eq:sm_phase_sum}
\end{align}    
with 
\begin{align}\label{eq:sm_relative_phase}
    \Phi_t(a, b) = \phi_t(a)+\phi_t(b)-\phi_t(\tilde{a})-\phi_t(\tilde{b}) \mod 2\pi.
\end{align}
The entropy growth is thus controlled by when the relative phase becomes nonzero.

\section{Two-species particle model}
\label{sec:sm_two_species}

In this section, we estimate the time evolution of the purity in the quantum automaton by describing the phase dynamics in Eq.~\eqref{eq:sm_phase_sum} in terms of a classical two-species exclusion process.

In the quantum automaton, the phases in Eq.~\eqref{eq:sm_relative_phase} are produced only by CZ gates, as we saw in Sec.~\ref{sec:sm_czswap_reduction}. If a CZ gate acts on the sites $(i,i+1)$ during the update from $t$ to $t+1$, then
\begin{align}\label{eq:sm_CZ_phase}
\Phi_{t+1}(a,b)-\Phi_t(a,b)
&=
\pi\Bigl[
a(t)_ia(t)_{i+1}
+b(t)_ib(t)_{i+1}
 -\tilde a(t)_i\tilde a(t)_{i+1}
-\tilde b(t)_i\tilde b(t)_{i+1}
\Bigr]
\mod 2\pi.
\end{align}
To track the time evolution of the phase, it is useful to collect the four time-evolved bits at each site into the vector
\begin{align}
\mathbf{s}(t)_i
=
\qty(a(t)_i,b(t)_i,\tilde a(t)_i,\tilde b(t)_i).
\end{align}
If $a(t)_i=b(t)_i$, then the two replicas have the same local bit before the replica swap, and their contribution to Eq.~\eqref{eq:sm_CZ_phase} is insensitive to exchanging the replicas. We therefore regard the corresponding site $i$ as empty. If $a(t)_i\neq b(t)_i$, the site is instead occupied by a particle. Following Refs.~\cite{Han2022,Han2023Jan,Han2023Feb}, we distinguish two particle species according to the corresponding bits in the original and swapped replicas:
\begin{align}
\mathbf{s}_i=(1,0,0,1)\ \mathrm{or}\ (0,1,1,0)
&\quad\Rightarrow\quad \mathrm{black},\\
\mathbf{s}_i=(1,0,1,0)\ \mathrm{or}\ (0,1,0,1)
&\quad\Rightarrow\quad \mathrm{white}.
\end{align}
At $t=0$, this rule has a simple geometric meaning. For $i\in A$, the replica swap exchanges the two copies,
\begin{align}
(\tilde a(0)_i,\tilde b(0)_i)
=
(b(0)_i,a(0)_i),
\qquad i\in A,
\label{eq:sm_initial_swap_in_A}
\end{align}
whereas for $i\in\bar A$ it does nothing,
\begin{align}
(\tilde a(0)_i,\tilde b(0)_i)
=
(a(0)_i,b(0)_i),
\qquad i\in\bar A.
\label{eq:sm_initial_swap_outside_A}
\end{align}
Thus a mismatch between the two replicas in a site inside $A$ is associated with a black particle, while a mismatch outside $A$ corresponds to a white particle. For example, if $A=\{1,2,3\}$ and
\begin{align}
a(0)=101010,
\qquad
b(0)=011000,
\end{align}
then
\begin{align}
\begin{array}{c|cccccc}
 &1&2&3&4&5&6\\
\hline
a(0)&1&0&1&0&1&0\\
b(0)&0&1&1&0&0&0\\
\tilde a(0)&0&1&1&0&1&0\\
\tilde b(0)&1&0&1&0&0&0\\
\hline
\mathrm{particle}&B&B&0&0&W&0
\end{array}
\label{eq:sm_particle_example}
\end{align}
Here $B$ and $W$ denote black and white particles. Notice that the third site is empty because the two bits $a_3$ and $b_3$ are equal, even though they are both $1$.

The two elementary gates in Eq.~\eqref{eq:sm_czswap_gate} have distinct roles in this particle language. A SWAP gate acting on sites $i$ and $i+1$ exchanges the corresponding bits in all four replicas, namely $a_i(t)\leftrightarrow a_{i+1}(t)$, $b_i(t)\leftrightarrow b_{i+1}(t)$, $\tilde{a}_i(t)\leftrightarrow \tilde{a}_{i+1}(t)$, and $\tilde{b}_i(t)\leftrightarrow \tilde{b}_{i+1}(t)$. Therefore, it simply exchanges the two four-bit vectors $\mathbf{s}_i$ and $\mathbf{s}_{i+1}$, transporting particles without changing their species. Since each site is associated with a single four-bit vector, at most one particle can occupy a single site. The action of the random SWAP gates thus corresponds to a two-species symmetric exclusion process for the black and white particles. A CZ gate does not change any computational-basis bit, so it does not move the particles. Its only effect is the phase in Eq.~\eqref{eq:sm_CZ_phase}, which depends on the two neighboring columns $\mathbf{s}_i$ and $\mathbf{s}_{i+1}$. For a same-species pair or a particle-empty pair, the four products in Eq.~\eqref{eq:sm_CZ_phase} cancel modulo $2$. For a neighboring black-white pair, they do not cancel, so the factor $e^{\im\Phi}$ changes sign. Specifically, substituting the possible particle configurations on the bond $(i,i+1)$ into Eq.~\eqref{eq:sm_CZ_phase} gives
\begin{align}
\begin{array}{c|c}
\text{particle configuration on }(i,i+1) & \Delta\Phi/\pi \mod 2\\
\hline
BB,\ WW,\ B0,\ 0B,\ W0,\ 0W,\ 00 & 0\\
BW,\ WB & 1
\end{array}.
\label{eq:sm_phase_table}
\end{align}
For instance, a $BW$ bond with
\begin{align}
\mathbf{s}_i=(1,0,0,1),
\qquad
\mathbf{s}_{i+1}=(1,0,1,0)
\end{align}
gives
\begin{align}
\Delta\Phi/\pi
=
1\cdot 1+0\cdot 0-0\cdot 1-1\cdot 0
=1
\mod 2,
\end{align}
whereas a $BB$ bond gives $1\cdot 1+0\cdot 0-0\cdot 0-1\cdot 1=0$, modulo $2$. For a particle-empty bond, one of the two sites has identical bits in the two replicas, and the same substitution gives $\Delta\Phi=0$, modulo $2\pi$.

We now use this table to reduce the phase sum in Eq.~\eqref{eq:sm_phase_sum} to a counting problem. Let $N_{\mathrm{nc}}(t)$ be the number of initial bit-string pairs $(a,b)\in\mathcal{C}^2$ for which black and white particles have never become adjacent up to time $t$, and define
\begin{align}
\mathcal{F}(t)
=
\frac{N_{\mathrm{nc}}(t)}{|\mathcal{C}|^2}.
\label{eq:sm_F_definition}
\end{align}
For these pairs, every CZ gate acts either on a same-species pair, on a particle-empty pair, or on two empty sites. According to Eq.~\eqref{eq:sm_phase_table}, each such gate gives $\Delta\Phi=0$ modulo $2\pi$. Since $\Phi=0$ at $t=0$, the phase remains zero throughout the evolution, and hence
\begin{align}
\sum_{\substack{(a,b)\in\mathcal{C}^2\\ \mathrm{no\ adjacency\ up\ to\ }t}}
e^{\im\Phi_t(a,b)}
=
N_{\mathrm{nc}}(t).
\label{eq:sm_no_contact_contribution}
\end{align}

For the remaining bit-string pairs $(a,b)$, there is at least one time step $t'\leq t$ at which a black and a white particle become adjacent. Once a black and a white particle become adjacent, the application of a CZ gate on that bond changes the sign of $e^{\im\Phi}$. Since, according to Eq.~\eqref{eq:sm_czswap_gate}, a given bond is acted on by a CZ gate with probability $1/2$ at each time step, these signs become effectively random after averaging over the random gates and the initial bit-string pairs. Consequently, these contributions effectively cancel upon averaging, leaving the coherent contribution in Eq.~\eqref{eq:sm_no_contact_contribution},
\begin{align}
P(t)
&=
\frac{1}{|\mathcal{C}|^2}
\sum_{(a,b)\in\mathcal{C}^2}
e^{\im\Phi_t(a,b)}
\simeq
\frac{N_{\mathrm{nc}}(t)}{|\mathcal{C}|^2}
=
\mathcal{F}(t).
\label{eq:sm_purity_F}
\end{align}
This is the relation used in the main text.

\section{Exclusion-process estimate of the purity}
\label{sec:sm_leading_particle}

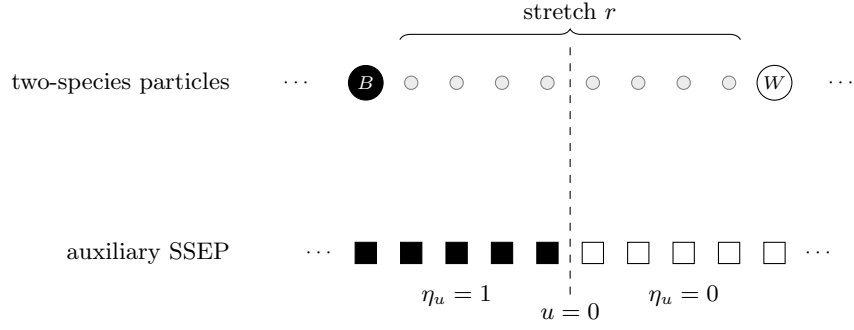
\begin{figure}
\centering
\begin{tikzpicture}[x=0.60cm,y=1.0cm]
\node[anchor=east] at (-0.8,2.25) {two-species particles};
\node[anchor=east] at (-0.8,0) {auxiliary SSEP};
\node at (0.5,2.25) {\scriptsize $\cdots$};
\foreach \x in {3,...,10} {
  \node[draw=gray,fill=gray!15,circle,inner sep=0pt,minimum size=5pt] at (\x,2.25) {};
}
\node[draw,circle,fill=black,text=white,inner sep=1.2pt,minimum size=13pt] at (2,2.25) {\scriptsize $B$};
\node[draw,circle,fill=white,text=black,inner sep=1.2pt,minimum size=13pt] at (11,2.25) {\scriptsize $W$};
\node at (12.5,2.25) {\scriptsize $\cdots$};
\draw[decorate,decoration={brace,amplitude=4pt}] (2.75,2.82) -- (10.25,2.82)
  node[midway,yshift=11pt] {stretch $r$};
\draw[dashed] (6.5,-0.55) -- (6.5,2.82);
\node[below] at (6.5,-0.55) {$u=0$};
\node at (1,0) {\scriptsize $\cdots$};
\foreach \x in {2,...,6} {
  \node[draw,fill=black,rectangle,inner sep=0pt,minimum size=8pt] at (\x,0) {};
}
\foreach \x in {7,...,11} {
  \node[draw,fill=white,rectangle,inner sep=0pt,minimum size=8pt] at (\x,0) {};
}
\node at (12,0) {\scriptsize $\cdots$};
\node[below] at (4,-0.32) { $\eta_u=1$};
\node[below] at (9,-0.32) {$\eta_u=0$};
\end{tikzpicture}
\caption{Auxiliary SSEP used to estimate the probability that black and white particles never become adjacent. The upper row shows a two-species configuration specified by the rightmost black particle and the leftmost white particle. The empty stretch between them has length $r$, shown here for $r=8$, and the dashed line is placed at its middle. The lower row shows the corresponding auxiliary SSEP on the same displayed positions. Filled and open squares denote occupied and empty sites in the auxiliary SSEP, corresponding to $\eta_u=1$ and $\eta_u=0$, respectively.}
\label{fig:sm_auxiliary_ssep}
\end{figure}

We now estimate the fraction $\mathcal{F}(t)$ defined in Eq.~\eqref{eq:sm_F_definition}. Since black particles are initially inside $A$ and white particles outside $A$, only the dynamics near the boundary of $A$ is relevant. For a fixed initial bit-string pair $(a,b)$, consider one such boundary. Let $x_B$ and 
$x_W$ denote the initial positions of the rightmost black particle inside $A$ and the leftmost white particle inside $\bar{A}$, respectively, see the upper panel of Fig.~\ref{fig:sm_auxiliary_ssep}. The initial empty stretch between them has length
\begin{align}
r
=
x_W-x_B-1.
\label{eq:sm_no_contact_condition}
\end{align}
Thus, a black-white contact occurs only when the initial empty stretch of length $r$ between them closes during the dynamics.

To determine the probability that the empty stretch is never closed during the dynamics, we introduce an auxiliary single-species SSEP, illustrated in Fig.~\ref{fig:sm_auxiliary_ssep}.
This auxiliary process is only a mathematical construction and does not represent an additional physical species. We replace the initial empty stretch between the rightmost black particle at $x_B$ and the leftmost white particle at $x_W$ by a domain-wall initial condition: the sites on one side of the center of the stretch are occupied, while the sites on the other side are empty. The boundary between these two regions defines the domain wall. This reference point is generally different from the subsystem boundary, which only specifies on which side the black and white particles are initially located. The coordinate of the auxiliary SSEP is not the original qubit coordinate. We denote it by $u$ and choose $u=0$ at the domain wall. In this auxiliary coordinate, the distance between the domain wall and the original rightmost black particle is $r/2$. Equivalently, this particle is identified with the $k=r/2$ particle of the auxiliary SSEP. 
Because particles of the same species obey exclusion and cannot overtake one another, the auxiliary particles can be labeled by their initial distance from the domain wall. Their evolution is therefore described by the SSEP with domain-wall initial data,
\begin{align}
\eta_u(0)
=
\begin{cases}
1,& u<0,\\
0,& u\geq 0,
\end{cases}
\label{eq:sm_domain_wall_ssep}
\end{align}
where $\eta_u=1$ means that the auxiliary site $u$ is occupied. To describe the motion of the leftmost white particle, we use the same auxiliary SSEP construction with the initial domain-wall configuration reversed. Let $x_k(t)$ be the distance reached by the $k$th auxiliary particle counted from the edge of the occupied domain. For the SSEP, its asymptotic position is~\cite{conroy2025}
\begin{align}
x_k(t)
\simeq
\sigma\sqrt{\frac{t}{\ln t}}
\qty[\ln t-\ln k].
\label{eq:sm_ssep_front}
\end{align}
Here $\sigma$ is defined by the diffusive scaling $\mathrm{Var}[x(t)]\simeq\sigma^2t$ of a tagged particle in the auxiliary SSEP. We use Eq.~\eqref{eq:sm_ssep_front} only to extract the leading $\sqrt{t\ln t}$ scale below.

For an initial empty stretch of length $r$, the relevant particles in the two auxiliary SSEPs are the $k=r/2$ particles counted from the two domain walls. As these particles evolve, they may reach each other and close the empty stretch. We define $r_*(t)$ as the characteristic initial length of the empty stretch below which the black and white particles typically become adjacent by time $t$. Thus, for $r<r_*(t)$ the corresponding particles typically meet before time $t$, whereas for $r>r_*(t)$ they typically do not. At the threshold $r=r^*(t)$, each particle has advanced by approximately $r^*(t)/2$, so the relevant particle label is $k\simeq r^*(t)/2$ when counted from the edge of the occupied domain. Hence $r^*(t)$ is determined, at leading order, by
\begin{align}
r_*(t)
\simeq
2x_{r_*(t)/2}(t).
\label{eq:sm_meeting_condition}
\end{align}
Solving Eq.~\eqref{eq:sm_meeting_condition} with Eq.~\eqref{eq:sm_ssep_front}, we obtain
\begin{align}
r_*(t)
\simeq
\sigma\sqrt{t\ln t},
\label{eq:sm_rstar}
\end{align}
at leading order in $t$.
Thus the bit-string pairs that contribute coherently to the purity are those whose initial empty stretch separating black and white particles is longer than $r_*(t)$.

For the initial state considered in the main text, the particle density in the two-species SSEP is
\begin{align}
n(m)=\frac{1-m}{2},
\label{eq:sm_particle_density}
\end{align}
where $m=M/N$ and $M$ is the number of the sites initialized in $\ket{0}$. Indeed, a site initially in $\ket*{0}$ is empty in the two-species SSEP because both replicas have the fixed value $0$. A site initially in $\ket*{+}$, instead, has independent bits in the two replicas and therefore hosts a particle whenever the two bits differ, which occurs with probability $1/2$. Thus the probability that a site is empty is $1-n(m)=(1+m)/2$. Equation~\eqref{eq:sm_purity_F} shows that the purity is controlled by the probability of finding an empty stretch longer than $r_*(t)$. For a stretch of length $r$, the two endpoint sites must be occupied by the rightmost black particle and the leftmost white particle, and the $r$ sites between them must be empty. The probability of an empty stretch of length $r$ is therefore $n(m)^2[1-n(m)]^r$. As a result, on average, 
\begin{align}
\mathbb{E}[\mathcal{F}(t)]
\simeq
\sum_{r>r_*(t)}
n(m)^2[1-n(m)]^r. 
\label{eq:sm_surviving_fraction}
\end{align}
Using $P(t)\simeq\mathcal{F}(t)$, the assumption $\mathbb{E}[\log_2\mathcal{F}(t)]\simeq \log_2 \mathbb{E}[\mathcal{F}(t)]$, and Eq.~\eqref{eq:sm_rstar}, we obtain
\begin{align}
\mathbb{E}[\log_2 P(t)]
\simeq
\sigma\sqrt{t\ln t}\log_2[1-n(m)],
\label{eq:sm_entropy_from_survival}
\end{align}
at leading order in $t$. 

We now evaluate $\sigma$ for the brickwork quantum automaton. We follow a tagged auxiliary SSEP particle over full brickwork time steps and write $x(t)=\sum_{s=0}^{t-1}\delta x_s$, where $\delta x_s$ is the displacement of the particle at each full time step. One full time step consists of two staggered SWAP layers. Since each SWAP is applied with probability $1/2$, the tagged particle has four possible displacements in one full time step:
\begin{align}
\delta x_s
&\in
\begin{cases}
\{+2,+1,0,-1\},& x_s\ {\rm even},\\
\{+1,0,-1,-2\},& x_s\ {\rm odd},
\end{cases}
\label{eq:sm_tagged_displacement}
\end{align}
where the four outcomes in each line occur with equal probabilities. The corresponding mean displacement is $+1/2$ for an even starting site and $-1/2$ for an odd starting site. After one full time step the parity is even or odd with equal probability, so this parity-dependent drift only gives an $O(1)$ correction to the displacement and does not affect the coefficient of the linear growth of $\mathrm{Var}[x(t)]$.
The one-step second moment is independent of the initial parity:
\begin{align}
\mathbb{E}[\delta x_s^2\mid x_s\ {\rm even}]
&=
\frac{1}{4}(2^2+1^2+0^2+(-1)^2)
=
\frac{3}{2},\\
\mathbb{E}[\delta x_s^2\mid x_s\ {\rm odd}]
&=
\frac{1}{4}(1^2+0^2+(-1)^2+(-2)^2)
=
\frac{3}{2}.
\label{eq:sm_tagged_second_moment}
\end{align}
Consecutive displacements are correlated because the displacement in one step determines the parity of the site from which the next step starts. Starting the next step from an even site gives a mean displacement of $+1/2$, whereas starting from an odd site gives a mean displacement of $-1/2$. Therefore,
\begin{align}
\mathbb{E}[\delta x_s\delta x_{s+1}\mid x_s\ {\rm even}]
&=
\frac{1}{4}
\qty[
2\cdot\frac{1}{2}
+1\cdot\qty(-\frac{1}{2})
+0\cdot\frac{1}{2}
+(-1)\cdot\qty(-\frac{1}{2})
]
=
\frac{1}{4},\\
\mathbb{E}[\delta x_s\delta x_{s+1}\mid x_s\ {\rm odd}]
&=
\frac{1}{4}
\qty[
1\cdot\frac{1}{2}
+0\cdot\qty(-\frac{1}{2})
+(-1)\cdot\frac{1}{2}
+(-2)\cdot\qty(-\frac{1}{2})
]
=
\frac{1}{4}.
\label{eq:sm_tagged_neighbor_correlation}
\end{align}
The correlations between displacements separated by two or more time steps vanish because the intermediate full time step randomizes the parity of the particle position. The displacement at the later time step therefore has zero mean:
\begin{align}
\mathbb{E}[\delta x_s\delta x_{s+r}]
=
0
\qquad (r\geq 2).
\label{eq:sm_tagged_long_correlation}
\end{align}
Putting these moments together gives
\begin{align}
\mathrm{Var}[x(t)]
&=
\sum_{s=0}^{t-1}\mathbb{E}[\delta x_s^2]
+2\sum_{s=0}^{t-2}\mathbb{E}[\delta x_s\delta x_{s+1}]
+O(1)\\
&=
\frac{3}{2}t
+2\cdot\frac{1}{4}(t-1)
+O(1)
=
2t+O(1).
\label{eq:sm_tagged_variance_result}
\end{align}
With the definition of $\sigma$ given below Eq.~\eqref{eq:sm_ssep_front}, Eq.~\eqref{eq:sm_tagged_variance_result} gives
\begin{align}
\sigma^2=2.
\label{eq:sm_sigma}
\end{align}
Combining Eqs.~\eqref{eq:sm_entropy_from_survival} and \eqref{eq:sm_sigma}, we obtain
\begin{align}
\mathbb{E}[\log_2 P(t)]
\simeq
\sqrt{2t\ln t}\log_2[1-n(m)].
\label{eq:sm_log_purity_final}
\end{align}
Using $S(\rho_A(t))=-\log_2P(t)$ and $1-n(m)=(1+m)/2$, and imposing that the entanglement entropy cannot be larger than $N_A(1-m)$, yields the entropy growth of the original Clifford circuit, because the dynamics of $|\mathcal{G}_A(t)|$ is the same as in the quantum automaton, as shown in Sec.~\ref{sec:sm_czswap_reduction}:
\begin{align}
\mathbb{E}[S(\rho_A(t))]
\simeq
\min\qty[
\sqrt{2t\ln t}
\log_2\qty(\frac{2}{1+m}),
N_A(1-m)
].
\label{eq:sm_entropy_final}
\end{align}
Using the relation between the entanglement asymmetry and the entanglement entropy derived in the main text, this gives
\begin{align}
\mathbb{E}[\Delta S(t)]
\simeq
\max\qty[
N_A(1-m)
-
\sqrt{2t\ln t}
\log_2\qty(\frac{2}{1+m}),
0
].
\label{eq:sm_asymmetry_final}
\end{align}
Eqs.~\eqref{eq:sm_entropy_final}~and~\eqref{eq:sm_asymmetry_final} constitute the two main analytic results of this work.

\section{Generality of the QME}
\label{sec:sm_general_initial_states}

In the previous section, we assumed the initial state in Eq.~\eqref{eq:ini_st_sm}. We now show that the same results apply to a generic product stabilizer state of the form
\begin{align}
\ket*{\Psi_{\mathrm{gen}}(0)}
=
\bigotimes_{i\in\Lambda_Z}\ket*{z_i}
\bigotimes_{j\in\bar{\Lambda}_Z}\ket*{\chi_j},
\label{eq:sm_general_initial_state}
\end{align}
where $\ket*{z_i}\in\{\ket*{0},\ket*{1}\}$ and $\ket*{\chi_j}\in\{\ket*{\pm},\ket*{\pm\im}\}$, with $\ket*{\pm\im}=(\ket*{0}\pm\im\ket*{1})/\sqrt{2}$. The set $\Lambda_Z$ has size $M=mN$. As before, we average over the choice of $\Lambda_Z$ at fixed $M$. Notice that, for the state~\eqref{eq:sm_general_initial_state}, $m$ is not necessarily the charge density, but rather the fraction of sites initialized in a computational-basis state. The initial state~\eqref{eq:ini_st_sm} used in the main text corresponds to the special case $\ket*{z_i}=\ket*{0}$ and $\ket*{\chi_j}=\ket*{+}$ for all sites.

We first make the stabilizer structure explicit. If we write
\begin{align}
\ket*{\chi_j}
=
\frac{\ket*{0}+e^{\im\varphi_j}\ket*{1}}{\sqrt{2}},
\qquad
\varphi_j\in\qty{0,\pi,\frac{\pi}{2},-\frac{\pi}{2}} ,
\label{eq:sm_general_local_states}
\end{align}
then the local stabilizer generators at $t=0$ are
\begin{align}
g_i(0)&=(-1)^{z_i}Z_i,
\qquad i\in\Lambda_Z,\\
h_j(0)&=\epsilon_j P_j,
\qquad j\in\bar{\Lambda}_Z,
\end{align}
where $P_j\in\{X_j,Y_j\}$ and $\epsilon_j=\pm1$. More explicitly,
\begin{align}
\begin{array}{c|cccc}
\ket*{\chi_j} & \ket*{+} & \ket*{-} & \ket*{+\im} & \ket*{-\im}\\
\hline
h_j(0) & X_j & -X_j & Y_j & -Y_j
\end{array}.
\label{eq:sm_general_xy_table}
\end{align}
Thus, the general initial state differs from the initial state~\eqref{eq:ini_st_sm} by the additional signs $(-1)^{z_i}$ and $\epsilon_j$, and by replacing some $X_j$ generators with $Y_j$ generators.

Under the symmetric Clifford circuit, or equivalently under the auxiliary quantum automaton, these generators evolve as
\begin{align}
g_i(t)
&=
(-1)^{z_i}Z_{r_i(t)},
\qquad i\in\Lambda_Z,
\label{eq:sm_general_z_evolution}\\
h_j(t)
&=
\epsilon_j P_{x_j(t)}
\prod_{\ell\in\Gamma_j(t)}Z_\ell,
\qquad
P_{x_j(t)}\in\{X_{x_j(t)},Y_{x_j(t)}\},
\qquad j\in\bar{\Lambda}_Z .
\label{eq:sm_general_non_z_evolution}
\end{align}
The stochastic variables $r_i(t)$, $x_j(t)$, and $\Gamma_j(t)$ are generated by the update rules in Eqs.~\eqref{eq:sm_Z_update} and \eqref{eq:sm_P_update}. These rules do not depend on $z_i$, $\epsilon_j$, or on whether the non-$Z$ operator is $X$ or $Y$.
The number of stabilizers supported entirely in $A$ is determined only by the variables  $r_i(t)$, $x_j(t)$, and $\Gamma_j(t)$. Multiplying a generator by single-site $Z$ generators can change the $Z$ factors in $\Gamma_j(t)$, but it cannot change the position $x_j(t)$ of its unique non-$Z$ operator. Thus, whether a generator can be supported entirely in $A$ is unaffected by the signs $(-1)^{z_i}$, $\epsilon_j$, or by whether the non-$Z$ operator is $X$ or $Y$. For a fixed realization of $r_i(t)$, $x_j(t)$, and $\Gamma_j(t)$, the number of stabilizers supported entirely in $A$ is therefore the same for the general initial state and for the reference initial state with $\ket{z_i}=\ket{0}$ and $\ket{\chi_j}=\ket{+}$. Hence,
\begin{equation}
|\mathcal{G}_A^{\mathrm{gen}}(t)|
=|\mathcal{G}_A^+(t)|,
\label{eq:sm_general_same_GA}
\end{equation}
where $|\mathcal{G}_A^{\mathrm{gen}}(t)|$ and $|\mathcal{G}_A^+(t)|$ are the numbers of independent stabilizer generators supported in $A$ for the general initial state~\eqref{eq:sm_general_initial_state} and for the initial state with $\ket*{z_i}=\ket*{0}$ and $\ket*{\chi_j}=\ket*{+}$, respectively. The equality therefore also holds after averaging over all the possible initial states with the same $M$ and over the dynamics.
Consequently, the results obtained through the two-species particle model for the initial states~\eqref{eq:ini_st_sm} can be directly applied here. In particular, the average entanglement entropy is
\begin{align}
\mathbb{E}_{\mathrm{gen}}[S(\rho_A(t))]
\simeq
\min\qty[
\sqrt{2t\ln t}
\log_2\qty(\frac{2}{1+m}),
N_A(1-m)
],
\label{eq:sm_general_entropy_final}
\end{align}
and the average entanglement asymmetry
\begin{align}
\mathbb{E}_{\mathrm{gen}}[\Delta S(t)]
\simeq
\max\qty[
N_A(1-m)
-
\sqrt{2t\ln t}
\log_2\qty(\frac{2}{1+m}),
0
].
\label{eq:sm_general_asymmetry_final}
\end{align}
They are the same expressions as in the main text, including the long-time saturation, with $m$ interpreted now as the fraction of sites initialized in a computational-basis state.  Therefore, the quantum Mpemba effect occurs for the initial states~\eqref{eq:sm_general_initial_state} by the same mechanism: decreasing $m$ increases the initial entanglement asymmetry, but it also increases the particle density $n(m)$, making black and white particles more likely to become adjacent. This causes the entanglement entropy to grow faster and, in turn, accelerates the decay of the entanglement asymmetry.

\end{document}